\documentstyle[preprint,eqsecnum,aps]{revtex}
\begin{document}
\draft
\preprint{}
\title{Gamma rays and neutrinos from a powerful cosmic accelerator}
\author{K. Mannheim$^{1,2}$}
\address{$^1$Max-Planck-Institut f\"ur Radioastronomie\\
Auf dem H\"ugel 69,  W--5300 Bonn 1\\ E-mail:  p036kma@mpifr-bonn.mpg.de
(INTERNET)\\
 $^2$ Universit\"ats--Sternwarte\\
Geismarlandstr. 11, D-37083 G\"ottingen\\
E-mail:  KMANNHE@gwdgv1.dnet.gwdg.de (INTERNET)\\
 Federal Rebublic of Germany}
\date{\today}
\maketitle
\begin{abstract}
Possibly, the powerful radio quasar 3C273 will reveal its nature
as an efficient proton accelerator up to energies of order $10^{11}$~GeV
in the near future.  It is shown in this paper that the shock accelerated
protons expected to be present in the quasar's plasma jet induce
an unsaturated synchrotron cascade with electromagnetic radiation
emerging in the X--ray and gamma ray range.  While (including the synchrotron
emission from the accelerated primary electrons) the broadband nonthermal
emission from 3C273 can be explained over the observed 18 orders of
magnitude, a flattening of the spectrum at the highest observed energies
(a few GeV) is predicted that could be falsified
by the Energetic Gamma Ray Experiment Telescope (EGRET) on board the
Compton Gamma Ray Observatory (GRO).  Above $\approx 100$~GeV the cascade
spectrum dramatically steepens again due to the absorption of the
gamma ray photons by the host galaxy's strong infrared photon field
from extended dust clouds -- in accordance with the non-detection of
3C273 by Cher\'enkov telescopes.  However, neutrinos from the hadronic
interactions initiating the cascade are not damped and reach terrestric
experiments without any modification of their injected flux.  In contrast
to the neutrino flux from pp--interactions, which are energetically unimportant
in jets, p$\gamma$--interactions generate a flat neutrino flux.
Therefore
it is emphasised that one must not simply normalize the expected
neutrino flux by the observed gamma ray flux.  Hence it is shown
that the expected neutrino flux in the energy range relevant for
underwater or underice detectors is much lower than assumed by many
authors.   On the other hand, with an increasing number of cosmic
gamma ray sources at known positions, their neutrino
detection should be feasible
when it is realized that angular resolution is
the crucial design property for neutrino detectors.
\end{abstract}

\pacs{95.30, 95.70.R, 98.70.S, 96.40.P}
\narrowtext

\section{Introduction}
\label{intro}

Recent gamma ray observations by GRO may be the clue to the
long--standing problem of the origin of UHE--cosmic ray protons.
GRO has discovered that many flat--spectrum radio quasars are strong
gamma ray sources with energy spectral index
$s=\log{[dN_\gamma/dE]}/d\log{[E]}$
above 100~MeV ranging between roughly
$-1.5$ and $-2.5$.  As pointed out below
the emission indicates the presence
of very energetic baryons that interact with the dense
photon atmosphere in the plasma outflow emerging
from the active galactic nucleus (AGN).
A prominent example is the nearby quasar 3C~273
with index $-2.39\pm 0.13$\cite{montigny92} which is investigated
in this paper because the steep gamma ray spectrum observed may
help to discriminate between the proton powered and other --
purely leptonic
-- models.

\subsection{The nonthermal ``mini-blazar''}
\label{intro:mini}

3C~273 is a well--known superluminal radio quasar at redshift $z=0.158$
({\it e.g.} Ref.\ \onlinecite{zensus88}).  VLBI
experiments showed that radio knots propagate
along
a strongly collimated jet emerging from a
stationary core with apparent
opening angle $\Phi_{\rm ob}=4^\circ$.  The motion of the knots with respect to
the core is
superluminal with
apparent speed $\beta_{\rm ob}=7h^{-1}$ over
a projected distance of at least $r_{\rm ob}=45h^{-1}$~pc at 5~GHz
where $h=H_\circ/100$~Mpc km$^{-1}$ s$^{-1}$.  The total flux of the
jet comprises the
flux of several knots with increasing self--absorption frequency
towards the base of the jet
and is flat ($\alpha\simeq 0$)
up to a spectral break in the mm range above
which the spectrum steepens ($\alpha\simeq 1$).
Detailed studies\cite{impey89}
of the variability and polarization
of 3C~273 have
disentangled the nonthermal
emission component associated to the compact jet
(the low--entropy ``miniblazar") from
the thermally emitting (high--entropy)
sites, such as the UV and soft
X--ray excess emitting ``cold" gas or the infrared emitting
dust.

The miniblazar can be attributed to a relativistic plasma outflow
from the AGN with embedded magnetic fields and shocks
\cite{blandford79}.  The shocks accelerate particles and the accelerated
electrons loose their energy rapidly due to observable
radio-optical synchrotron emission in the
magnetic field.
Interferometric radio maps rather suggest a helix
pattern for the trajectories of the radio knots and, indeed,
the situation can be pictured as a topological
equivalent to the curved shocks in the solar wind following the
Parker spiral of the magnetic field -- with the equatorial plane
of the sun being bent onto a cone, so that the spiral becomes a helix,
and the shock waves being excited by mass ejections from the accreting
gas at the base of the jet.
Remarkably, the accelerated protons and nuclei \cite{biermann87}
were long thought to be negligible as radiative agents, since they
have much lower energy losses at the same energy as electrons.
However, acceleration thus operates longer for the heavy particles
until their energy losses also become large.   The energy lost
by the energetic baryons then emerges as cascade radiation in the
X-gamma ray band.

Since the shock speed does not reflect the fluid speed (in which
the scattering centers move with the Alfv\'en speed), the speeds
of knots may be different, {\it i.e.} a strong reconfinement shock
\cite{sanders83}
after an initial passage of a thick cloud may be a standing or a
weakly--relativistic shock wave.  As a consequence the emission from
this shock, particularly the infrared/X--ray emission, could be less
beamed than the radio emission,
since radio
emission is self--absorbed within the inner parsec of a jet and thus
comes from further out.
Beaming
of the emitted radiation pattern results from the anisotropy
arising from the Lorentz transformation between the comoving fluid
frame (with isotropic particle and photon distributions)
and the observer's frame
(``Doppler--boosting'').  However,
for the sake of simplicity it is assumed in this paper that
the Doppler factor is the same along the jet.
\subsection{UHE--protons}
\label{intro:protons}

The crucial question is, of course, wether
protons can really reach the extremely
high energies  needed to generate a luminosity of
secondaries competitive
with inverse--Compton or synchrotron--self--Compton emission from the
accelerated electrons and to be able to provide enough
energy in the center--of--momentum frame of the
colliding protons and photons to create secondary
particles on the mass shell.

The maximum possible energy is readily obtained from
\begin{equation}
t_{\rm p,cool}(E_{\rm p,max})=t_{\rm acc}(E_{\rm p,max})
\label{emax1}\end{equation}
provided that the limit from drift across the shock with radius $r_\perp$
\begin{equation}
E_{\rm p,max}\approx  e({\bf v\times B}){\bf r}/c\simeq
7.8\cdot 10^{20}
\beta_{\rm s}\left(B\over {\rm G}\right)\left(r_\perp\over {\rm pc}\right)
\ {\rm eV}\label{emax2}\end{equation}
\cite{eichler81}  is not exceeded.
Here $t_{\rm i}$ denotes the time scale of process i, $B$ the magnetic
field and $\beta_{\rm s}$ the speed of the shock.
Due to energy losses the proton distribution steepens from
$dN/dE\propto E^{-2}$ (or even flatter, cf.
Ref.\ \onlinecite{ellison90}) to $dN/dE\propto E^{-3}$
when
\begin{equation}
t_{\rm p,cool}(E_{\rm p,b})=t_{\rm e,cool}(E_{\rm e,b})=
t_{\rm exp}\label{ebreak1}\end{equation}
where $t_{\rm exp}$ denotes the dynamical time scale of the expanding jet.
When $E_{\rm p,b}\ge E_{\rm p,max}$, no
steepening occurs and the proton distribution turns over with a
Bessel--function
type behaviour at $E_{\rm p,max}$.  Since
\begin{equation}
{L_{\rm p}\over L_{\rm e}}\simeq
{u_{\rm p}\over u_{\rm e}}{t_{\rm e,cool}\over t_{\rm p,cool}}\simeq
\eta {{\rm Min}\left[E_{\rm p,b},
E_{\rm p,max}\right]\over E_{\rm p,b}}\label{lple}
\end{equation}
a small $E_{\rm p,max}$ costs a great value of $\eta=u_{\rm p}/u_{\rm e}$
to obtain comparable luminosities from protons and electrons, respectively.
Note that $\eta$ in Eq. (\ref{lple}) refers to the energy density ratio
at extremely relativistic energies, so that the Galactic value
$\eta(>\rm GeV)\approx 100$ indeed lets one expect the possibility that
$L_\gamma\gg L_{\rm ir}$.
With $B\propto r^{-1}$
along the jet of length $r$
and $r$ small enough, so that $E_{\rm p,b}\le
E_{\rm p,max}$,
it follows that
$L_{\rm p}/L_{\rm e}\propto \eta$, whereas
further out $L_{\rm p}/L_{\rm e}\ll \eta$.
Therefore, one expects to observe
gamma rays mainly from the compact
regions of radio jets close to the core.  Additionally,
the emission from the nuclear jet (the blazar) is
concentrated in a narrow lighthouse beam due to
relativistic bulk motion of the plasma amplifying
the flux at small angles to the line of sight.
{\it Thus it
can be explained,  why flat--spectrum radio
sources (indicating jet sources at small angles
to the l.o.s.) are the gamma ray sources detected
in flux--limited experiments}.

Comparing 3C~279 \cite{mannheim93} and 3C~273 one can see that in spite of a
sim
ilar
inferred radiation compactness, the maximum proton energy in the
jet of 3C~273 is smaller
because of the presence of the  strong external radiation field
seen as the big blue bump.
For the maximum dimensionless energy it follows
from Eq.\ (\ref{ebreak1}) that
\begin{equation}
\gamma_{\rm p,b}=\left(m_{\rm p}\over m_{\rm e}\right)^3
{1+a_{\rm s}\over 1+\langle \sigma_{\rm p\gamma}/
\sigma_{\rm p,syn}\rangle
a_{\rm s}}\gamma_{\rm e,b}
\label{ebreak2}\end{equation}
with the target photon/magnetic energy density ratio
$a_{\rm s}=u_\gamma/u_B$ and the factor
$\langle \sigma_{\rm p\gamma}/
\sigma_{\rm p,syn}\rangle\simeq 240$ accounting for
the relative importance of photoproduction and
synchrotron energy losses, respectively.
  For the intrinsic
photons $a_{\rm s,in}\approx \beta_{\rm j}\gamma_{\rm j}
\Phi/(1+\eta)\ll 1$, whereas
the external
photons with luminosity $L_{\rm uv}\simeq 10^{47}$~erg/s
enhance the local photon density
such that
$\alpha_{\rm s,ex}=2L_{\rm uv}/(r_{\rm b}+r_{\rm c})^2
B_{\rm b}^2c\simeq 0.1$ assuming $r_{\rm b}=r_{\rm c}$
for the length $r_{\rm b}$ of the radiative jet and
the distance $r_{\rm c}$
of the base of the jet to the location
of the thermal photons ({\it i.e.} collimation out to
the parsec scale).
The value thus obtained for the proton break energy
$\gamma_{\rm p,b}=3\cdot 10^{10}$ (Tab. \ref{table})
is in modest agreement with the value for $\gamma_{\rm p,max}$
obtained from
Eq.\ (\ref{emax1}).  Thus, head--on
collisions with a (infrared) photon of energy
$\varepsilon_{\rm t}= 145\thinspace{\rm MeV}/2
\gamma_{\rm p,b}\simeq2\cdot 10^{-3}$~eV can
excite the $\Delta$ resonance.  Note that the observed
energies are cosmologically redshifted and Doppler
blueshifted yielding $E_{\rm p,max}({\rm obs})=
D_{\rm j}\gamma_{\rm p,max}/(1+z)\simeq 2\cdot 10^{11}$~
GeV, if there were no energy loss during propagation
through the cosmic microwave background.

Since the pair absorption cross section peaks for
head--on collisions, but the external photons
appear
highly anisotropic in the comoving fluid frame,
{\it  the
pair creation
opacity ($\propto$ compactness) in the very forward
direction remains unaffected by the additional target
photons}.  In contrast, the Thomson scattered
luminosity is found to be
$L_{\rm T}\approx a_{\rm s,ex}L_{\rm s}$,
where $L_{\rm s}$ denotes the apparent soft synchrotron
luminosity of the jet, since the Thomson scattering is isotropic in the
comoving
 frame.
In models where the particle acceleration takes place
much closer to the AGN, {\it i.e.} $r_{\rm c}\ll 1$~pc,
the thermal seed photon
energy density could exceed the magnetic field energy
density in
the jet, so that, indeed, the Compton--upscattered
thermal photon flux can dominate over the emission from
the jet itself
(cf. Ref.\ \onlinecite{melia89} or
for another version of the same model see Ref.
\onlinecite{dermer92}).
Very close to the AGN the extremely
strong Compton cooling
is faster than diffusive acceleration processes which
are rather slow for reasonable diffusion coefficients.
Only explosive acceleration seems possible in this case.
On the other hand the well--ordered structure of the
jet on much larger scales must not be destroyed by
such processes, nor must there be too efficient
dissipation, for otherwise the luminosity of
the large scale jet could not be as large as observed.

In the following Sections it is shown how
the cascade respondes to a varying proton maximum
energy and how the infrared photons from dust clouds
modify the emerging spectrum.  Finally, the neutrino
flux from 3C273 is calculated.

\section{The unsaturated synchrotron cascade and the few MeV
bump}
\label{cascade}

It is instructive to calculate step by step the development
of the unsaturated synchrotron cascade induced by the
UHE protons.
For each step conservation of the
total emitted
power $L\propto x^2 dN_\gamma / dx$ determines the normalization of the
stationa
ry
photon spectrum $dN_\gamma/dx$.
The result of a numerical
calculation is shown in Fig.\ \ref{figure}
for the physical conditions listed in
Tab. \ref{table} and the detailed integral equation solved is discussed
in \cite{mannheim91}.

We start with protons of Lorentz factor $\gamma_{\rm p,b}=3\cdot 10^{10}$
producing pions of Lorentz factor $\gamma_\pi=
\left(m_{\rm p}/m_{\pi}\right)
\gamma_{\rm p,b}/5$.  The neutral pions decay yielding two gamma rays
of dimensionless energy
\begin{equation}
x_\circ={E_\gamma\over m_{\rm e}{\rm c}^2}\simeq {m_{\rm p}\over m_{\rm e}}
{1\over 10}\gamma_{\rm p,b}\simeq 180\gamma_{\rm
p,b}\label{xmax0}\end{equation}
in the comoving frame
(multiplication with the redshifted Doppler factor
$D_{\rm j}/(1+z)$ yields the observed frequencies or
energies).
The local {\it emissivity} of gamma rays is $Q_\gamma\propto x^{-1}$ and
the {\it stationary photon spectrum} steepens
to $dN_\gamma /dx\propto x^{-2}$ because of
the $\gamma\gamma$ pair production losses.
Here we take
$dN_{\rm p}/d\gamma_{\rm p}\propto \gamma_{\rm p}^{-2}$ (shock acceleration,
neglibible losses below $\gamma_{\rm p,b}$)
and $dN_{\gamma,\rm target}/dx\propto x^{-2}$ (as observed between $10^{11}$~Hz
and $10^{15}$~Hz corresponding to proton energies at threshold between
$\gamma_{\rm 1}=m_\pi/(2m_{\rm e}x_1)=10^{12}$ and $\gamma_2=10^8$, resp.).
The reason for the flat injected flux of secondary gamma rays
is the following:
Compared to cooling on monoenergetic target photons with energy
$x_{\rm m}$, which yield $Q_\gamma\propto x^{-2}$ above
$\gamma_{\rm m}=m_\pi/(2m_{\rm e}x_{\rm m})$, cooling in the inverse power
law photon distribution with $N_{\gamma,\rm target}(>x)\propto x^{-1}$ yields
additional target photons ($x<x_{\rm m}$) for
$\gamma_{\rm p}>\gamma_{\rm m}$
and hence a flatter emissivity
$Q_\gamma\propto x^{-2+1}$ \cite{mannheim89}.
Now, we forget about
charged pions, direct pair production and proton
synchrotron
emission, because
the pairs from charged pion decay simply add more
flux to the neutral pion decay cascade, direct pairs radiate
mostly around TeV where -- as will be shown--
the emission is absorbed by dust infrared
emission and the proton synchrotron emission
contributes only for $a_{\rm s}< 0.004$.

The first synchrotron photon generation is now radiated by the pairs
produced from the original neutral pion decay quanta via
\[\gamma +\gamma_{\rm t}\rightarrow e^++e^-\]
where $\gamma_{\rm t}$ denotes a soft photon at threshold energy
$xx_{\rm t}=1$ (head--on collision).
The maximum electron Lorentz factor is given by $\gamma_1=x_\circ/2$
and the characteristic synchrotron frequency is then
\begin{equation}
x_1=3\cdot 10^{-14}B_\perp\gamma_1^2=2.2\cdot 10^{11}B_\perp\label{
xmax1}\end{equation}
The stationary electron distribution being subjected to synchroton
losses is $dN_{\rm e}d\gamma_{\rm e}\propto \gamma^{-2}$ (for $\gamma
<\gamma_1$
)
and therefore the synchrotron
emissivity is $Q_\gamma\propto x^{-3/2}$ (for $x<x_1$).
As long as we are in the
optical thick range, the stationary photon spectrum is steeper by one
power, {\it i.e.} $dN_\gamma /dx\propto x^{-5/2}$.
The pairs thus produced have Lorentz factors $\gamma_2=x_1/2$
and radiate the second synchrotron generation
\begin{equation}
x_2=3\cdot 10^{-14}B_\perp \gamma_2^2=3.6\cdot 10^8 B_\perp^3
\label{xmax2}\end{equation}
with stationary photon spectrum $dN_\gamma /dx\propto x^{-11/4}$.
However, part of this generation is already at energies below
\begin{equation}
x^*\simeq {10\over l}=5\cdot 10^6\label{xstern}\end{equation}
where the jet becomes optically thin (compactness in the comoving frame
$l=2\cdot 10^{-6}$ for 3C~273. cf. Tab.1), and therefore has a spectrum
$dN_\gamma /dx\propto x^{-7/4}$.
The energy $x^{**}$ where the pairs produced at $x^*$ radiate
is given by
\begin{equation}
x^{**}=3\cdot 10^{-14}B_\perp \left[x^*\over 2\right]^2
=0.2B_\perp\label{xsternstern}\end{equation}
The third generation has the characteristic energy
\begin{equation}
x_3=3\cdot 10^{-14}B_\perp \gamma_3^2= 10^3B_\perp^7
\label{xmax3}\end{equation}
Now, here comes the most important point:  There are two special
energies involved.  The energy $x^*$ defined by $\tau_{\gamma\gamma}
(x^*)=1$
and the energy $x_2$, which may be either
$<,=$ or $>$ than $x^*$.  In the first case the second cascade
generation is optically thin with respect to pair creation
(no further reprocessing) and has
a spectrum with
index $-7/4$.  In
the second and in the third case, there will be a third cascade
generation.  The part of the spectrum from the optically thick range
$x_2>x>x^*$ (which is
monoenergetic for $x_2=x^*$) has the form
$dN_\gamma /dx\propto x^{-15/8}$ over an energy band $\Delta \log\left[x
\right]=2\log\left[
x_2/x^*\right]$ (a l\'a 3C~279) from $x^{**}$ to $x_3$
and the part from the optically thin
range has the spectrum $dN_\gamma /dx\propto x^{-11/8}$ emerging below
$x^{**}$.  Thus, $\alpha_\gamma\simeq 0.9$ and $\alpha_X\simeq 0.4$
ignoring that the superposition of
individual cascade generations somewhat smears out these values.

The case $x^*\approx x_2$, relevant for 3C~273 where $x_2$ follows the
reduction of $\gamma_{\rm p,b}$, leads to a gap between the maximum of the
third generation at $x^{**}\approx x_3$ and the maximum of the second
(partially
absorbed) generation at $x^*$.  It is not possible to predict
the accurate shape of this gap, because it depends sensitively
on the upper
turnover of the injection {\it modulo} all decay kinematics and
successive synchrotron smearing ($x\propto \gamma^2$).  The numerical
result (assuming $B_\perp\approx B_{\rm b}$ so that $D_{\rm j}x_3(1+z)\approx
2$, that is 1~MeV)
with injection of an exponential turnover is shown in Fig.1
for 3C~273 yielding an effective spectral index close to the
observed $-2.39$.  However, at a few GeV the spectrum of the
decadic power rises again due to the onset of the second cascade
generation with photon index $-7/4$ up to 10~TeV$\approx D_{\rm j}x^*/
(1+z)$.

The energy $x_3\propto B_\perp^7$ is
extremely dependent upon the magnetic field strength
in contrast to $x^{**}\propto B_\perp$.  Therefore one may expect
a wide range of spectra in the MeV--GeV range.  There is no way,
however, to understand a few MeV bump without an accompanied intrinsic bump
at 1--10~TeV, since the cascade generation preceding the one
that makes the MeV bump must show up at $D_{\rm j}x^*/
(1+z)$.

\section{Infrared absorption of the few TeV bump}
\label{infrared}

This brings us to the final argument, {\it viz.} the absorption
of gamma rays outside of the jet  by photons from the
extended regions of 3C~273.

The optical depth for scattering on a monoenergetic
photon target is approximately given by
\begin{equation}
\tau_{\gamma\gamma}(x)=r_{\rm ext}
N_\gamma(x_\circ)\sigma(x)\label{tau1}\end{equation}
where $x=h\nu/m_{\rm e}c^2$ denotes the dimensionless gamma ray photon energy,
$\sigma(x)\simeq {3\over 16}\sigma_{\rm T}
x_{\rm th}/x$ for $x\ge x_{\rm th}=2/[x_\circ(1-\cos\theta)]$ is the pair
creation cross section above threshold energy $x_{\rm th}$, $\sigma_{\rm T}$
the Thomson cross section and
$N_\gamma(x_\circ)=L_\circ
/(4\pi r_{ext}^2m_{\rm e}c^3x_\circ)$ the external target photon
density of the source with
monochromatic luminosity $L_\circ$ at target
energy $x_\circ$.
For $L_\circ=10^{46}L_{46}$~erg/s
and $x\ge x_{\rm th}$ this yields
\begin{equation}
\tau_{\gamma\gamma}(x)=
4\cdot 10^{15}L_{46}(r_{\rm ext}/{\rm cm})^{-1}x_{\rm th} x_\circ^{-1} x^{-1}
\label{tau2}\end{equation}
so that at threshold $\tau_{\gamma\gamma}=4\cdot
10^{15}L_{46}(r_{\rm ext}/{\rm cm})^{-1}x_\circ^{-1}$.
Assuming that the infrared emission comes from  dust at
$r_{\rm ext}=300$~pc we obtain that $\tau_{\gamma\gamma}\ge 1$ for
$x\ge 5\cdot 10^5$ corresponding to $E_\gamma\ge 100$~GeV.
Since the produced pairs would isotropize rapidly, the reemitted
radiation in the m.f. of the host galaxy does not remember the beam,
so that the  apparent luminosity is reduced to the comoving frame
luminosity which is down by a large kinematic factor.

One can further ask, wether the flattening of the predicted
cascade spectrum above 10~GeV
(cf. Fig.~\ref{figure}) could be destroyed by
photon--photon absorption as well as the 10~TeV bump.
The absorbing
thermal optical/X--ray photons presumably come
from behind the jet (from the apex) and therefore the angle $\theta$
should have the value obtained for the angle between
jet axis and observer, {\it viz.}
$\theta=7^\circ$.  With $x$ denoting the observed gamma ray energy,
the resonant target photon energy is $x_\circ\simeq 1.4\cdot 10^{-2}$
(7~keV).
The optical depth at threshold for $L_{46}(7\rm keV)=0.1$ is given by
$\tau_{\gamma\gamma}(r_{\rm c}+r_{\rm b}=8
\thinspace{\rm pc}, \theta=7^\circ)\simeq 6\cdot 10^{-4}$;
clearly insufficient to steepen the spectrum.

\section{Conclusions}
\label{conclusions}

Radio jets emerge out of the vicinity of an accreting compact object,
presumably a black hole, where a rotating magnetosphere tied
to the accreting gas feeds a considerable fraction of the infalling
matter into the outflow, thereby transporting angular momentum outwards.
The magnetic fields allow for collimated flow solutions \cite{camenzind90}
passing through bulk
Lorentz factors of 5-20 beyond the lightcylinder at roughly
$10r_{\rm G}\simeq 1.5\cdot 10^{15}m_9$~cm, where $m_9$ denotes
the mass of the black hole in units of $10^9m_\odot$.
If the initial acceleration of plasma would be much more efficient, {\it i.e.}
$\gamma_{\rm i}\approx  10^3$, the ambient
thermal UV and soft X--ray photons from the accreted gas
excert a `Compton drag'
on the particles in the
outflow
slowing it down to terminal Lorentz factors $\gamma_\infty\approx 10$,
thus dissipating most of its energy into
Compton--scattered radiation \cite{melia89}.
However, this produces maximum photon
energies well below $100$~MeV -- in apparent contradiction with the
observed fluxes {\it above} GeV from many extragalactic radio sources.
Moreover, if radio jets consist of ordinary hydrogen plasma,
radio--loud objects like 3C~273 have a very powerful jet even
many kiloparsecs away from the AGN and therefore it seems unlikely that
they have dissipated much of their power in the central parsec
already.

After some inital collimation by toroidal fields the jet
expands freely and super--Alfv\'enically into the surrounding medium.
However, free expansion tends to bring the internal pressure of
the jet within a few scale heights above the Alfv\'en point
into equilibrium with the external pressure.
The flow tries to realise an equilibrium state by reconfinement
shocks \cite{sanders83}, which let the internal pressure vary in
a zig--zag fashion about the external pressure.  Thus, the external
gas determines the structure of the jet and the amount of radiative
dissipation, which establishes itself through shock acceleration
at the reconfinement edges.  Another important source of shocks could
be mass ejection at the base of the jet -- much like in the solar wind.
Thus, when the ejection sites are tied to a rotating accretion disk, a helical
shock pattern results, which seems to be required by VLBI observations of
many radio jets
and HST observations of the M87 jet.

Synchrotron emission of accelerated electrons emerges
mostly in the radio to optical regime, with the submm/infrared/optical
emission coming from the innermost nuclear jet ($r\le r_{\rm b}\approx
4$~pc) and the
radio emission coming from further out.  Note, however,
that the propagation of the jet
through the central parsec
(out to the BLR) is assumed to be essentially
non--dissipative, {\it i.e.} $r_{\rm c}=O[1{\rm pc}]$.
During this collimated stage the initial Poynting
flux is converted into bulk kinetic flux, which is
then the reservoir for particle acceleration at
shocks farther out.
Higher frequencies than optical
are difficult to achieve from
synchrotron emission by the primary
accelerated electrons, because the maximum energies are
strongly constrained by energy losses \cite{biermann87}.
However, the protons accelerated in the jet reach energies so high
($\gamma_{\rm p,max}\approx 10^{10}$) that
saturation effects due to the finite size of the shocks become
important, simply because their cooling time scale, which
is dominated by secondary particle production for a wide parameter
range, is very long
compared to the cooling time scale of electrons.
Thermal photons from outside of the jet can additionally
damp the protons in the jet.
The power induced by the cooling ultra--high--energy protons
at comoving frame energies up to $10^{10}$~GeV is rapidly reprocessed by an
electromagnetic shower and emerges in the X-- to gamma ray
regime.  The infrared to optical emission from the throat of
the jet serves as the scattering agent for the cascade determining
the photon energy $\varepsilon_\gamma\approx 10 $~TeV
where the jet becomes optically thick with
respect to pair creation in the observer's frame.

When the initial jet collimation terminates
very far away from the
central engine or when the thermal emission from the central engine is
inherently weak, then the emerging cascade spectrum is much like as
it is observed by GINGA, COMPTEL and EGRET for
3C~279:  $\alpha_X\approx 0.6$ with
steepening at a few MeV, so that $\alpha_\gamma\approx 1.0$.  On
the other hand, a quasar like 3C~273 harbouring a ``miniblazar'', but
otherwise emitting predominantly thermal UV and IR emission, is
very likely to show signs of interactions of these photons with the
accelerated particles inside the nuclear jet.  The interactions reduce
the maximum energies of the particles they can obtain by diffusive
acceleration limited by energy losses.  As shown, this reduction moves
the injection energy of secondaries at the top of the cascade closer
to the energy where the jet becomes optically thick with respect to
pair creation.
Therefore 3C~273 developes a
different cascade spectrum with two bumps at a few MeV and TeV, respectively.
The latter is absorbed by infrared photons from dust clouds within the host
galaxy, whereas the MeV bump has power law wings on both sides in energy
space matching the observed X--ray and gamma ray power--laws consistent
with GINGA, COMPTEL and EGRET observations.  Of course, this does not
rule out the existence of additional emission components in the X--ray
band.

For 3C~273
flattening of the spectrum
at a few GeV is expected, but observations do not seem to indicate this --
a detail deserving further measurements.
It is at the present stage of investigation unclear, what the variability
properties should be like.  Variability must be very complex,
because we are actually dealing with a source extending over scales
from fractions of a parsec to parsecs (transverse size of the jet or
downstream emission regions of shocks).
However, the variability time scale of the submm--nearinfrared
target photons $t_{\rm ir}\simeq r_{\rm b,\perp}(1+z)/D_{\rm j}c\simeq
180$~hours)
should at least be contained in the X-- and gamma ray
autocorrelation functions, the amplitude should be significantly
larger because of cascade reprocessing.
An outburst in the power of the jet itself should manifest itself
first by a flaring
submm/infrared/optical continuum (the target) followed very
shortly after by the gamma rays
and then much later by a corresponding rise
of the radio flux.  Since the radio flux peaks at its self--absorption
frequency, the delay
at a given radio frequency
determines the scale length of the radio emitting part of
the jet, {\it viz.}  $\Delta t\simeq (r-r_{\rm b})(1+z)/D_{\rm j}c$
with $r=4(\nu_{\rm obs}/\nu_{\rm b})^{-1}
\thinspace{\rm pc}$ and $\nu_{\rm b}=2\cdot 10^{10}$~Hz.

As a corollary of the interpretation of the gamma rays
from 3C~273 being of hadronic origin it follows that
3C~273 should also be a source, which
provides roughly {\bf 1\%} of {\bf a)}
the flux of cosmic rays at $10^{19}$~eV at Earth and
{\bf b)}
the atmospheric background neutrino flux at 10~TeV
in a solid angle of
$1^\circ \times  1^\circ$.
To come to these
conclusions consider the following:

(i)  In contrast to the electromagnetic flux
the neutron as well as the neutrino flux
density spectrum from photomeson production
are flat (the luminosity increases proportional
with the energy), because the UHE protons in the jet scatter
on a polychromatic target, {\it i.e.} an inverse power law,
so that the number of target photons increases rapidly with
increasing energy.   The electromagnetic cascade washes out
the injection energy of the photons
and redistributes the power below the energy
above which the jet is optically thick with respect to
pair creation.

(ii)  Hadronic
proton--photon interactions produce neutrons by isospin flip.
The {\it apparent} neutron luminosity $L_{\rm n}\approx L_{\pi^+}$
is given by $L_{\rm n}\approx
{4\over 13}L_\gamma$ (Ref.\ \onlinecite{mannheim92}
where pion and pair production
are taken into account) and the maximum
observed energy is
$E_{\rm n}={1\over 2}D_{\rm j}E_{\rm p,max}/(1+z)$.  Hence one
obtains using $\int_x^\gamma F_\nu d\nu=3\cdot 10^{13}$~Jy~Hz
\begin{equation}
F_{\rm n\rightarrow p}={4\over 13}E_{\rm n,max}^{-1}\int_x^\gamma
F_\nu d\nu\simeq 7\cdot 10^{-19}\thinspace{\rm cm^{-2} s^{-1}}
\label{fneu}\end{equation}
while the observed flux at $10^{19}$~eV is roughly $6\cdot
10^{-17}$~cm$^{-2}$~s$^{-1}$ \cite{berezinski90}.
The neutrons
are not confined to the magnetic field in the jet and
because of the $\beta$--decay length
$l_\beta=100[\gamma_{\rm n}/10^{10}]$~kpc
they can escape the host galaxy without adiabatic losses
(cf. original papers in Ref.\ \onlinecite{berezinski90}).
Energy losses on the cosmic microwave background further reduce
the flux above $10^{19}$~eV.
Moreover, diffusion of the
protons due to random magnetic fields in the surroundings
of 3C~273 could smear the cosmic ray beam, thus reducing
the cosmic ray flux by a factor $D_{\rm j}^{-3}\simeq
3\cdot 10^{-3}$ and the particle energy by
$D_{\rm j}^{-1}=1/7$.  Such smearing would, however, increase the
total number of radio sources contributing to the entire
extragalactic flux, since then also non--beamed (not core dominated,
steep spectrum) sources would contribute (cf. Ref.\ \onlinecite{rachen93} for a
detailed
model with direct proton escape from isotropic hot spots
at the tip of powerful jets).  .

(iii)  The neutrino luminosity for all species is given by
$L_\nu\approx {3\over 13} L_\gamma$  and
$E_{\rm \nu,max}={1\over 20}D_{\rm j}E_{\rm p,max}/(1+z)$, see Ref.\
\onlinecite{mannheim92}. This yields
the flux
\begin{equation}
F_\nu(p\gamma)={3\over 13}E_{\rm \nu,max}^{-1}\int_x^\gamma F_\nu d\nu\simeq
5\cdot 10^{-18}\thinspace{\rm cm^{-2}s^{-1}}\label{fpgamma}\end{equation}
important at Fly's Eye energies \cite{mannheimstanev92}.
The neutrino flux
from $pp$--interactions (which dominates
at energies in the 10~TeV range, see Ref.\ \onlinecite{mannheim93})
is given by
\begin{equation}
F_\nu(pp)\simeq {10r_{\rm G}\over r_{\rm b}\ln{E_{\rm p,max}/E_{\rm p,min}}}
F_\nu(p\gamma)\left(E\over E_{\rm \nu,max}\right)^{-1}
\simeq 2\cdot 10^{-5}m_9F_\nu(p\gamma)
\left(E\over 10^{10}\thinspace{\rm GeV}\right)^{-1}\label{fpp}\end{equation}
reaching $10^{-16}m_9$~cm$^{-2}$~s$^{-1}$ at 10~TeV.  Thus,
the neutrino flux from the jet of 3C~273
is far below the background
neutrino flux from AGN
as predicted
by several authors (see Refs.\ \onlinecite{stenger92,stecker91}).  However,
the effective intensity for a solid angle of $1^\circ\times 1^\circ$
reaches
$3\cdot 10^{-17}m_9$~cm$^{-2}$~s$^{-1}$~sr$^{-1}$~GeV$^{-1}$
at 10~TeV
for 3C~273
alone, compared to $\approx 10^{-14}$~cm$^{-2}$~s$^{-1}$~sr$^{-1}$~GeV$^{-1}$
for the hypothetical AGN background (which itself becomes comparable with
the atmospheric background at 10~TeV).
Considering that the number of similar flat--spectrum
radio sources is very large \cite{kuehr81}
detection of the entire population of
these powerful {\it cosmic accelerators} seems possible
by refined techniques using horizontal atmospheric
showers \cite{halzen92} and by taking advantage from the fact
that the candidate source positions are known to any desired accuracy.

\acknowledgments
I wish to acknowledge support by DARA grant FKZ50OR9202.


%
%
\begin{figure}
\caption{The broadband spectrum of 3C~273 from
submm wavelengths up to hard gamma rays decomposed
into a ``blazar" component from a nuclear jet
and thermal emission from accreted gas.  Data for the
submm-optical blazar component are taken from Ref.\
\protect\onlinecite{impey89}, who
suggested the decomposition based upon polarization and
variability, for the X--ray range from Ref.\ \protect\onlinecite{staubert92},
for the hard X--ray/soft gamma ray range
from Ref.\ \protect\onlinecite{hermsen92}
and, finally, for the gamma
ray range from Ref.\ \protect\onlinecite{montigny92}.
Several attempts to detect 3C~273 with Cher\'enkov
telescopes have failed \protect\cite{weekes88}.
The synthetic spectrum shown comprises the emission
from shock accelerated electrons (submm--optical)
and protons (X--gamma rays).  The emission mechanism
is optically thin synchrotron emission from ``primary"
(=shock accelerated) electrons and from pairs produced
by an unsaturated synchrotron cascade initiated by
the protons cooling in the soft synchrotron photon soup.
The TeV emission is absorbed by infrared photons from
dust within the host galaxy.  The model predicts a
flattening of the spectrum between 10~GeV and a few hundred GeV.}
\label{figure}
\end{figure}
%

\begin{table}
\caption{
Physical conditions in the compact jet of 3C~273.
Basic parameters for the calculation of physical conditions
are: redshift $z$, Lorentz factor of jet $\gamma_{\rm j}$,
angle to the l.o.s. $\theta$, proton/electron energy ratio $\eta=L_\gamma/
L_{ir/o}$, equipartition constant $k_{\rm e}\approx u_{rel}/u_B$, luminosity
of jet in m.f. and rel. particles $L_{44}$ in units of $10^{44}$~erg/s,
observed opening angle of jet $\Phi_{\rm ob}$ and $\nu_{\rm c}$ comoving
frame cutoff frequency of the synchrotron spectrum from
the primary electrons.
Inferred quantities (consistent with observations)
are the observed apparent speed $\beta_{\rm ob}$,
the Doppler factor of the jet $D_{\rm j}=
[\gamma_{\rm j}(1-\beta_{\rm j}\cos\theta)]^{-1}$, the primary electron break
frequency $\nu_{\rm b}$, the photon/magnetic energy ratio $a_{\rm s}$
due to the radiation produced in the jet and due to an external luminosity
of $10^{47}$~erg/s, resp., the projected length of the
radiative jet $r_{\rm b,ob}$ where most of the emission with
$\nu>\nu_{\rm b}$ comes from,
the deprojected length $r_{\rm b}$, the (unresolved) transverse size of
the jet $r_\perp$, the target photon compactness in the comoving frame $l$,
the magnetic field $B_{\rm b}$ at $r_{\rm b}$, the electron
and proton break
Lorentz factors $\gamma_{\rm e,b}$ and $\gamma_{\rm p,b}$ and finally
the observed photon energy $E_\gamma^*$
where the jet becomes optically thick with
respect to pair creation.}
\label{table}
\begin{tabular}{cccc}
&\bf Basic parameters&\bf 3C~273 (miniblazar)&\bf  Comment, reference\\
\tableline
&$z$&$0.158$&\ \cite{kuehr81}\\
&$\gamma_{\rm j}$& ($\ge$) $7$&\ \cite{zensus88}\\
&$\theta$\rm [deg]&($\ge$) $8$&\ \cite{zensus88}\\
&$\eta$&$15$&$L_\gamma/L_{ir}$ (this work)\\
&$k_{\rm e}$&$1$&$u_{\rm rel}\approx u_B$ assumed
(this work)\\
&$L_{44}$&($\ge$) $50$&\ \cite{meisenheimer89} -- hot spot A\\
&$\Phi_{\rm ob}$~[deg]&$4$&\ \cite{zensus88}\\
&$\nu_{\rm c}$ [\rm Hz]&$ 10^{14}$&\ \cite{impey89}, cf.
\cite{biermann87}\\
\tableline
&\bf Inferred quantities&($\Lambda_{\rm e}=5$, $\Delta=5$)&\bf Equation\\
\tableline
&$\beta_{\rm ob}$&$7$&\ \cite{blandford79} Eq.~(1)\\
&$D_{\rm j}$& $7$&\ \cite{blandford79} Eq.~(6)\\
&$\nu_{\rm b}$ [\rm Hz]&$ 2\cdot 10^{10}$&\ \cite{blandford79}
Eq.~(30)\\
&$a_{\rm s,in}$&$ 0.004 $&\ \cite{blandford79} Eq.~(24), intrinsic\\
&$a_{\rm s,ex}$&$ 0.125 $&$\approx 2L_{\rm uv}/B_{\rm b}^2(r_{\rm b}
+r_{\rm c})^2c$\\
&$r_{\rm b,ob}$\rm [pc]&$0.6$&\ \cite{blandford79} Eq.~(31)\\
&$r_{\rm b}$\rm [pc]&$4$&$r_{\rm b,ob}/\sin\theta$\\
&$r_{\rm b,\perp}$\rm [pc]&$0.04$&$r_{\rm b,ob}\Phi_{\rm ob}$\\
&$l$&$2 \cdot 10^{-6}$&\ \cite{mannheim93} Eq.~(18)\\
&$B_{\rm b}$\rm [G]&$ 0.3 $&\ \cite{mannheim93} Eq.~(24)\\
&$\gamma_{\rm e,b}$&$1.5\cdot 10^{2}$ &\ \cite{mannheim93} Eq.~(7)\\
&$\gamma_{\rm p,b}$&$3\cdot 10^{10}$&\ \cite{mannheim93} Eq.~(15)\\
&$E_\gamma^*$(obs)\rm  [TeV]&$ 18 $&\ \cite{mannheim93} Eq.~(4)\\
\end{tabular}
\end{table}
\end{document}